# Quantum cryptography with polarizing interferometers

Marek Czachor

*Katedra Fizyki Teoretycznej i Metod Matematycznych*
*Politechnika Gdańska, ul. Narutowicza 11/12, 80-952 Gdańsk, Poland*

Cryptographic scheme proposed by Bennett, Brassard, and Mermin [Phys. Rev. Lett. **68**, 557 (1992)] is reformulated in a version involving two polarizing Mach-Zehnder interferometers. Such a form, although physically equivalent to the original one, makes its security explicit, suggestive and easy to explain to non-experts.

In the Ekert cryptographic scheme (E91) [1] pairs of EPR particles were used to generate identical sequences of bits in remote places, while Bell's theorem certified that the particles were not measured in transit by an eavesdropper. The modification of the Ekert scheme proposed by Bennett, Brassard and Mermin (BBM92) [2] did not use the Bell theorem and was based essentially on the fact that two particles correlated by a singlet state behave similarly to a single particle. A user of the cryptographic channel, traditionally called Alice or Bob, instead of sending a polarized particle through the channel, may make a measurement of polarization of his or her particle and somehow "create" an appropriately polarized object at the other side of the channel. In this way the BBM92 protocol becomes a kind of the original Bennett-Brassard 1984 (BB84) protocol [3] but with "polarization at a distance" [4].

Quantum cryptography based on entangled states is an ingenious practical application of the set of ideas and techniques that were originally developed in order to understand the problem of completeness and limitations of quantum mechanics. Another such application, equally amazing and simultaneously very simple, is the idea of interaction-free measuerements proposed by Elitzur and Vaidman (EV) [5]. Below I will discuss the latter in some detail since the objective of this Letter is to show that combining the ideas of BBM92 and EV one arrives at a particularly suggestive version of a quantum cryptographic scheme. The scheme can be formally shown to be secure by a direct application of the proof given in BBM92. What is interesting, however, its security is so *explicit* that an explanation of the problem to non-experts becomes particularly simple. As opposed to the standard schemes one can illustrate the point without referring to technical aspects such as complementarity, non-cloning theorem, or Bell's theorem with all its loopholes.

Let us begin with the EV experiment. It is based on the Mach-Zehnder interferometer consisting of two *identical* semi-transparent symmetric mirrors (Fig. 1). The action of such mirrors can be described by $2\times 2$ unitary maps

$$|\psi\rangle_{\rm in} \mapsto U|\psi\rangle_{\rm in} = |\psi\rangle_{\rm out}. \qquad (1)$$

Using a matrix notation one can write $U$ as a $2\times 2$ unitary matrix, say,

$$\begin{pmatrix} \psi_{1\,{\rm out}} \\ \psi_{2\,{\rm out}} \end{pmatrix} = \frac{1}{\sqrt{2}} \begin{pmatrix} i & 1 \\ 1 & i \end{pmatrix} \begin{pmatrix} \psi_{1\,{\rm in}} \\ \psi_{1\,{\rm in}} \end{pmatrix}. \qquad (2)$$

An essential ingredient of the interferometer is the $\pi/2$ phase shift occuring between the transmitted beams: For $\psi_{1\,{\rm in}} = 1$, $\psi_{2\,{\rm in}} = 0$ (i.e. $|\psi\rangle_{\rm in} = |1\rangle_{\rm in}$) one finds

$$U|1\rangle_{\rm in} = \frac{1}{\sqrt{2}}\Big(i|1\rangle_{\rm out} + |2\rangle_{\rm out}\Big). \qquad (3)$$

The fact that the two mirrors are identical implies that the twice-transmitted beam is shifted by $\pi$ with respect to the beam which is twice-reflected. And vice versa: The beam that has been reflected at the first mirror and then transmitted at the second one, has the same phase as the beam that has been first transmitted and then reflected. An interference between the upper and the lower paths results finally in a completely destructive interference in the "1-out" port and a completely constructive interference in the "2-out" port. Formally this looks as follows

$$U^2|1\rangle_{\rm in} = i|2\rangle_{\rm out} \qquad (4)$$

or $U^2 = i\sigma_x$. Now, without going into formal details, it is clear that had we closed one of the paths in the interferometer, the interference pattern would be destroyed and, independently of the choice of $|\psi\rangle_{\rm in}$, the output probabilities would be $1/2$. In this way a detection in the "1-out" port of a single photon may signal a presence of an obstacle in one of the paths even though there is no transfer of energy to the obstacle. This is the essence of the EV interaction-free measurement.

The cryptographic scheme I will describe below is based on a similar property of a *polarizing* Mach-Zehnder interferometer (Fig. 1a). A detailed theory of such a device was presented in [6] and the configuration of the cryptographic channel was inspired by the beautiful interferometric version of the EPR problem proposed by Żukowski and Pykacz [7].

The interferometer we will concentrate on is called polarizing because the first semi-transparent mirror is replaced by an analyzer of circular polarizations (denoted $U_\pm$): A right-handed beam incoming through the "1-in" port is reflected and goes "up", whereas a left-handed beam goes through the device unchanged. To get a close analogy with the EV problem we assume that the phase shifts between the two beams that leave the analyzer are identical to those generated by $U$ [compare (3)]:

$$U_{\pm}\Big(A|1_+\rangle_{\text{in}} + B|1_-\rangle_{\text{in}}\Big) = iA|1_+\rangle_{\text{out}} + B|2_-\rangle_{\text{out}}. \quad (5)$$

It is clear that a recombination of the two beams will not lead to any interference between the two paths of the interferometer because the two beams have different circular polarizations. Therefore before we recombine the two beams at a semi-transparent mirror, we first turn the left-handed beam into a right-handed one by means of a half-wave plate which is placed at the lower path of the interferometer. Such an interferometer can be used to make an interaction-free measurement if one takes a *linearly* polarized state (polarization along an $x$-axis)

$$|\psi\rangle_{\text{in}} = \frac{1}{\sqrt{2}}\Big(|1_-\rangle_{\text{in}} + |1_+\rangle_{\text{in}}\Big) = |1_x\rangle_{\text{in}} \quad (6)$$

as the input. Indeed, as we shall see in a moment, the entire interferometer maps $|1_x\rangle_{\text{in}}$ into $i|2_+\rangle_{\text{out}}$. Analogously, a $y$-polarized state

$$|1_y\rangle_{\text{in}} = \frac{1}{\sqrt{2}}\Big(|1_-\rangle_{\text{in}} - |1_+\rangle_{\text{in}}\Big) \quad (7)$$

is transformed into $|1_+\rangle_{\text{out}}$ (note that all these states are right-handed). The one-to-one relationship between the linear polarization states of the input and the appropriate "out" ports will be lost if some obstacle will close one path of the interferometer: An obstacle at the lower path removes a left-handed component from the incoming beam; an obstacle at the upper path removes a right-handed component. The effect of the obstacle is therefore equivalent to putting a circular polarization filter at the input of the interferometer. I point out this element to show that the one-to-one relationship we have just mentioned is physically equivalent to the EV effect. However, this is not what we are interested in here. What is important, the "EV effect" turns our polarizing interferometer into an analyzer of linear polarizations, a property of crucial importance for the cryptographic procedure.

Before we proceed further let us first make our discussion more formal. The Hilbert space of the input is now 4-dimensional (two output ports and two polarization states in each port makes four "out" states; by unitarity we must have also four "in" states). It is instructive to consider an interferometer with an arbitrary phase shift $\alpha$. A state will be written in the circular polarisation basis:

$$|\psi\rangle = \begin{pmatrix} \psi_{1_+} \\ \psi_{1_-} \\ \psi_{2_+} \\ \psi_{2_-} \end{pmatrix}. \quad (8)$$

The four parts of the interferometer are

$$U_{\pm} = \begin{pmatrix} i & 0 & 0 & 0 \\ 0 & 0 & 1 & 0 \\ 0 & 0 & 0 & 1 \\ 0 & 1 & 0 & 0 \end{pmatrix} \quad \text{(polarizing beam splitter)} \quad (9)$$

$$U_{\lambda/2} = \begin{pmatrix} 1 & 0 & 0 & 0 \\ 0 & 1 & 0 & 0 \\ 0 & 0 & 0 & 1 \\ 0 & 0 & 1 & 0 \end{pmatrix} \quad \text{(half wave plate)} \quad (10)$$

$$U_{\alpha} = \begin{pmatrix} e^{i\alpha} & 0 & 0 & 0 \\ 0 & e^{i\alpha} & 0 & 0 \\ 0 & 0 & 1 & 0 \\ 0 & 0 & 0 & 1 \end{pmatrix} \quad \text{(phase shifter)} \quad (11)$$

$$U = \frac{1}{\sqrt{2}} \begin{pmatrix} i & 0 & 1 & 0 \\ 0 & i & 0 & 1 \\ 1 & 0 & i & 0 \\ 0 & 1 & 0 & i \end{pmatrix} \quad \text{(symmetric mirror)} \quad (12)$$

and the entire interferometer is

$$V_{\alpha} = U U_{\alpha} U_{\lambda/2} U_{\pm} = \frac{1}{\sqrt{2}} \begin{pmatrix} -e^{i\alpha} & 1 & 0 & 0 \\ 0 & 0 & ie^{i\alpha} & 1 \\ ie^{i\alpha} & i & 0 & 0 \\ 0 & 0 & e^{i\alpha} & i \end{pmatrix}. \quad (13)$$

The fact that $V_{\alpha}$ is an analyzer of linear polarizations can be seen in the formulas

$$V_{\alpha}^{\dagger}|1_+\rangle_{\text{out}} = \frac{-1}{\sqrt{2}}\Big(e^{-i\alpha}|1_+\rangle_{\text{in}} - |1_-\rangle_{\text{in}}\Big) \quad (14)$$

$$V_{\alpha}^{\dagger}|1_-\rangle_{\text{out}} = \frac{-i}{\sqrt{2}}\Big(e^{-i\alpha}|2_+\rangle_{\text{in}} + i|2_-\rangle_{\text{in}}\Big) \quad (15)$$

$$V_{\alpha}^{\dagger}|2_+\rangle_{\text{out}} = \frac{-i}{\sqrt{2}}\Big(e^{-i\alpha}|1_+\rangle_{\text{in}} + |1_-\rangle_{\text{in}}\Big) \quad (16)$$

$$V_{\alpha}^{\dagger}|2_-\rangle_{\text{out}} = \frac{1}{\sqrt{2}}\Big(e^{-i\alpha}|2_+\rangle_{\text{in}} - i|2_-\rangle_{\text{in}}\Big). \quad (17)$$

Of special importance for us are the folmulas (14) and (16) since they explicitly show the linear polarization basis analyzed by $V_{\alpha}$. We shall rewrite these formulas as a definition of the new basis:

$$i|2_+\rangle_{\text{out}} = V_{\alpha}|\alpha\rangle_{\text{in}} \quad (18)$$

$$-|1_+\rangle_{\text{out}} = V_{\alpha}|\alpha_{\perp}\rangle_{\text{in}}. \quad (19)$$

In particular, $|0\rangle_{\text{in}} = |1_x\rangle_{\text{in}}$ and $|0_{\perp}\rangle_{\text{in}} = -|1_y\rangle_{\text{in}}$. Let us recall that the interferometer acts as the analyzer of *linear* polarizations as a result of the interference between the two *circular* polarizations: For any linearly polarized input state its different circular components propagate by different paths and interfere due to the presence of the half-wave plate.

Consider now a source producing either of the following two entangled states (Fig. 2).

$$|\Psi_{\text{in}}^{\pm}\rangle = \frac{1}{\sqrt{2}}\Big(|1_+\rangle_{\text{in}}|1_-\rangle_{\text{in}} \pm |1_-\rangle_{\text{in}}|1_+\rangle_{\text{in}}\Big) \quad (20)$$

where the circular polarization states correspond to photons propagating to the "left" and to the "right", respectively. Switching to the linear polarization basis we find that

$$|\Psi_{\text{in}}^+\rangle = \frac{e^{i\alpha}}{\sqrt{2}}\Big(|\alpha\rangle_{\text{in}}|\alpha\rangle_{\text{in}} - |\alpha_\perp\rangle_{\text{in}}|\alpha_\perp\rangle_{\text{in}}\Big) \quad (21)$$

$$|\Psi_{\text{in}}^-\rangle = \frac{e^{i\alpha}}{\sqrt{2}}\Big(|\alpha_\perp\rangle_{\text{in}}|\alpha\rangle_{\text{in}} - |\alpha\rangle_{\text{in}}|\alpha_\perp\rangle_{\text{in}}\Big) \quad (22)$$

and, since the formula is valid for any $\alpha$, we obtain two EPR-type states with entangled linear polarizations (parallel for $|\Psi_{\text{in}}^+\rangle$ and perpendicular for $|\Psi_{\text{in}}^-\rangle$). On the other hand, both states have maximally entangled circular polarizations. The entanglement of linear polarizations together with the analyzer-of-linear polarizations property of $V_\alpha$ explain the origin of violation of the Bell inequality in a coincidence experiment with the state $V_\alpha \otimes V_\beta |\Psi_{\text{in}}^\pm\rangle$. To give an example, one finds

$$\langle\Psi_{\text{in}}^+|V_\alpha^\dagger \otimes V_\beta^\dagger|P(1_+)_{\text{out}} \otimes P(1_+)_{\text{out}}|V_\alpha \otimes V_\beta|\Psi_{\text{in}}^+\rangle$$
$$= \langle\Psi_{\text{in}}^+|V_\alpha^\dagger \otimes V_\beta^\dagger|P(2_+)_{\text{out}} \otimes P(2_+)_{\text{out}}|V_\alpha \otimes V_\beta|\Psi_{\text{in}}^+\rangle$$
$$= \frac{1}{2}\cos^2\frac{\alpha-\beta}{2} \quad (23)$$

where the $P$'s are projectors corresponding to the outputs of the two interferometers. In particular, for $\alpha = \beta$ one obtains the perfect coincidence of detections. It is easy to understand this effect when we recall that for $|\Psi_{\text{in}}^+\rangle$ the linear polarizations of both photons are the same independently of the choice of the polarization plane and that the phase of the interferometer determines the plane of polarization analyzed by the interferometer. However, in a realistic experiment the result of the form (23) will hold approximately since for either $\alpha$ or $\beta$ nonzero the above formulas are valid for purely monochromatic fields. An efficiency will be maximal only for $\alpha = \beta = 0$. The above arrangement is therefore naturally suited for the EPR-type cryptography discussed by Bennett in [4] where the polarizations involved are linear-$x$, linear-$y$, left-handed and right-handed.

Assume Alice and Bob choose interferometers with no phase shifts (i.e. with equal lengths of the paths, similarly to the EV experiment, Fig. 1b) and a source producing

$$|\Psi_{\text{in}}^+\rangle = \frac{1}{\sqrt{2}}\Big(|1_x\rangle_{\text{in}}|1_x\rangle_{\text{in}} - |1_y\rangle_{\text{in}}|1_y\rangle_{\text{in}}\Big) \quad (24)$$

$$= \frac{1}{\sqrt{2}}\Big(|1_+\rangle_{\text{in}}|1_-\rangle_{\text{in}} + |1_-\rangle_{\text{in}}|1_+\rangle_{\text{in}}\Big) \quad (25)$$

and consider the four experiments symbolically denoted by

$$V_0 \otimes V_0 \quad (26)$$
$$V_0 \otimes U_\pm \quad (27)$$
$$U_\pm \otimes V_0 \quad (28)$$
$$U_\pm \otimes U_\pm \quad (29)$$

In (26) Alice and Bob both measure linear polarizations; in (27) Alice measures a linear polarization but Bob measures a circular one, and so on. Each of them chooses between $V_0$ and $U_\pm$ at random. The cryptographic key is produced, for example, when they both measure circular polarizations [i.e. (29)] but the test for eavesdropping is performed by comparison of data collected in (26) and a part of those from (29).

It is easy to show that the scheme is secure. Indeed, Eve in order to get an information about the key has to measure circular polarizations. After a measurement she sends to Alice and Bob appropriately polarized photons, say, $|1_-\rangle_{\text{in}}|1_+\rangle_{\text{in}}$. But Alice and Bob test for eavesdropping by measuring linear polarizations. The perfect correlations in the $V_0 \otimes V_0$ experiment are due to the "EV effect", that is are present if and only if each of the photons goes by *both paths in the interferometer*. However, whenever Eve sends photons that are circularly polarized, the photons propagate by only *one path* in the interferometers (a similar effect would be obtained if Eve had closed the upper path in Alice's interferometer and the lower path in Bob's one). The reason is obvious: A right-handed photon is with certainty reflected by $U_\pm$ and goes by the upper path, and the left-handed one is with certainty transmitted by $U_\pm$ and goes through the lower path. There is absolutely no way of sending a photon that with certainty would go by only one path and simultaneously by both paths. However, a part of data for measurements of circular polarizations must be also checked for coincidence to avoid the following rather nasty attack. Eve measures circular polarizations for all pairs, but then sends to Alice and Bob identical linear polarized photons. They check for eavesdropping and do not discover anything. Alice encodes the secret information according to the one-time pad procedure, but when it turns out that Bob cannot decode the information it is already too late since Eve knows the message. When they sacrifice a part of the possible key data they very quickly discover there is no perfect correlation in circular polarizations.

To formally prove the security one simply repeats the steps from BBM92 and substitutes circular polarizations for measurements of $\sigma_z$ and the linear ones for those of $\sigma_x$.

The procedure I propose is equivalent to BBM92, E91, and BB84. However, I believe it is much more intuitive and easier to explain to those who are not professionally involved in investigation of foundations of quantum mechanics. It clearly shows the analyzer-of-linear polarizations property of the polarizing Mach-Zehnder interferometer. The latter feature is very paradoxical in itself and can be used to prove a simple Bell-type theorem for a single photon [6].

[1] A. Ekert, Phys. Rev. Lett. **67**, 661 (1991).
[2] C. H. Bennett, G. Brassard, and N. D. Mermin, Phys. Rev. Lett. **68**, 557 (1992).
[3] C. H. Bennett and G. Brassard, in *Proceedings of IEEE International Conference on Computers, Systems, and Signal Processing, Bangalore, India* (IEEE, New York, 1984), p. 175.
[4] C. H. Bennett, Phys. Rev. Lett. **68**, 3121 (1992).
[5] A. Elitzur and L. Vaidman, Found. Phys. **23**, 987 (1993).
[6] M. Czachor, Phys. Rev. A **49**, 2231 (1994).
[7] M. Żukowski and J. Pykacz, Phys. Lett. A **127**, 1 (1988).

FIG. 1. (a) Polarizing Mach-Zehnder interferometer. $U_\pm$ is an analyzer of circular polarizations, $U$ a symmetric lossless beam splitter, $\lambda/2$ a half-wave plate, and $\beta$ a phase shifter. The upper paths are labelled "1" and the lower ones "2". $U_\pm$ is fully transparent for left-handed photons and fully reflecting for the right-handed ones if they arrive through the "1-in" port. The properties of the analyzer with respect to the photons arriving through the "2-in" port are irrelevant from the viewpoint of our discussion. (b) Ordinary Mach-Zehnder interferometer used in the Elitzur-Vaidman experiment. The beam of light arriving through the "1-in" port leaves through the "2-out" port.

FIG. 2. An experiment $V_\alpha \otimes V_\beta$. Alice (left) and Bob (right) measure linear polarizations of photons arriving in a singlet state. The polarization planes are determined by the phases of the interferometers. The experiment $U_\pm \otimes U_\pm$ corresponds to the situation where instead if recombining the beams at $U$ the detectors are placed directly behind $U_\pm$.

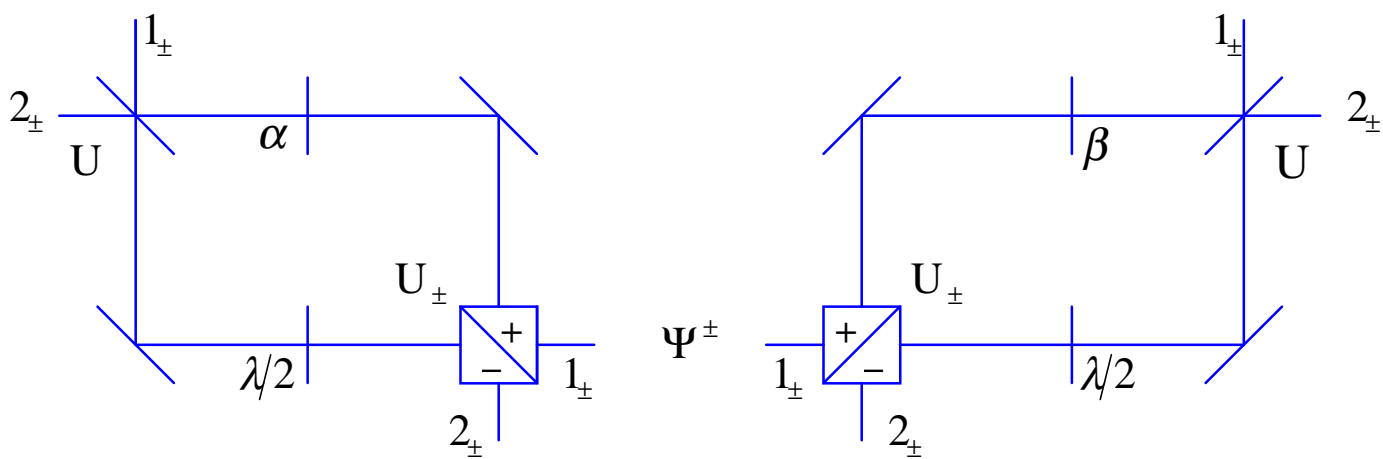

**Fig.2**

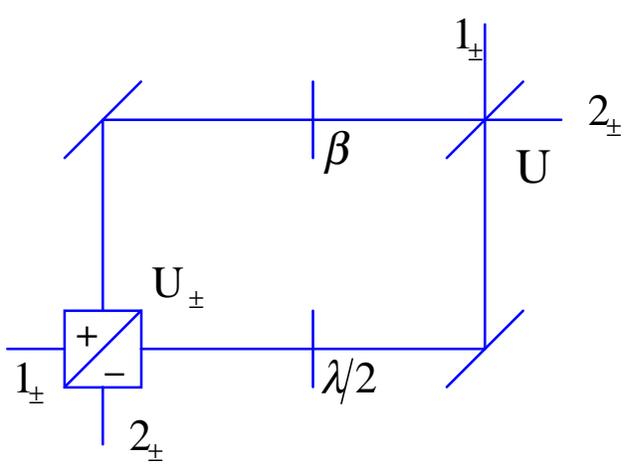
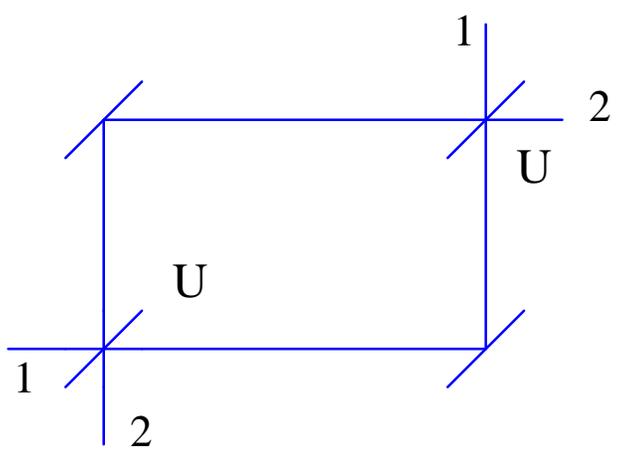

**Fig.1a**                                                            **Fig.1b**